\documentclass[aps,prb,reprint,longbibliography, superscriptaddresses]{revtex4-2}
\usepackage{mathrsfs}
\usepackage{textcomp, gensymb}
\usepackage{amsmath,gensymb}
\usepackage{amsfonts}
\usepackage{amssymb}
\usepackage{amsthm}
\usepackage{graphicx}
\usepackage{natbib}
\usepackage{color}
\usepackage{hyperref}
\usepackage{bm}
\usepackage[caption=false]{subfig}
\usepackage{verbatim}
\usepackage{braket}
\usepackage{physics}
\usepackage{nicefrac}
\usepackage{soul}

\newcommand{\alrcom}[1]{{\color{black} #1}}

\begin{document}
	\title{Fluctuations of the inverted magnetic state and how to sense them}

	\author{Anna-Luisa E. R\"omling$^1$}
 
        \author{Artim L. Bassant$^2$}
	\email{a.l.bassant@uu.nl}
        \author{Rembert A. Duine$^{2,3}$}
        \affiliation{$^1$Condensed Matter Physics Center (IFIMAC) and Departamento de F\'isica
			Te\'orica de la Materia Condensada, Universidad Aut\'onoma de Madrid,
				E-28049 Madrid, Spain \\
        $^2$Institute for Theoretical Physics, Utrecht University, Utrecht, The Netherlands \\
        $^3$Department of Applied Physics, Eindhoven University of Technology, P.O. Box 513, 5600 MB Eindhoven, The Netherlands 
        }

	\begin{abstract}
			Magnons are the low-energy excitations of magnetically ordered materials. While the magnetic moment of a ferromagnet aligns with an applied magnetic field, it has been experimentally shown that the magnetic order can be inverted by injecting spin current into the magnet. This results in an energetically unstable but dynamically stabilized state where the magnetic moment aligns antiparallel to an applied magnetic field, called the inverted magnetic state. The excitations on top of such a state have negative energy and are called antimagnons. The inverted state is subject to fluctuations, in particular, as shot noise in the spin current, which are different from fluctuations in equilibrium, especially at low temperatures. Here, we theoretically study the fluctuations of the inverted magnetic state and their signatures in experimental setups. We find that the fluctuations from the injection of spin current play a large role. In the quantum regime, the inverted magnetic state exhibits larger fluctuations compared to the equilibrium position, which can be probed using a qubit. Our results advance the understanding of the fundamental properties of antimagnons and their experimental controllability, and they pave the way for applications in spintronics and magnonics, such as spin wave amplification and entanglement.
	\end{abstract}
	\maketitle
	
	\section{Introduction}
        
        Magnonics concerns itself with the excitation, propagation, and manipulation of spin waves.
        A spin wave is an excitation in an ordered magnetic medium, and its quantum is known as a magnon.
        Although spin waves or magnons were introduced in 1932 by Bloch \cite{bloch_zur_1932}, only in the last 35 years have they become the subject of extensive research \cite{flebus_2024_2024}.
        They promise to be key to high-speed information processing and the future of computing.
        Several hindrances are yet to be dealt with before magnons are fully integrated into device applications.
        In particular, the number of magnons is not a conserved quantity because of their limited lifetimes.
        Many efforts have been put into investigating materials and devices to negate magnon damping.
        For example, Yttrium Iron Garnet (YIG) has proven to exhibit an extremely low damping coefficient \cite{hauser_yttrium_2016}.
        One can also use spin-orbit torque to inject angular momentum in the form of spin current, which extends the magnon's lifetime \cite{nikitchenko_spin-orbit_2021}.

        Furthermore, magnons may also prove to be useful for quantum technology.      
        In recent years, significant progress has been made in the entanglement of quasi-particles and quantum devices that utilize magnons.
        This emergent field is called quantum magnonics \cite{YUAN20221}, and it aims to combine the fields of magnonics and quantum information.
        However, entangling two distant magnons has been a challenge.
        Some proposals suggest using a cavity to entangle the magnetic degree of freedom \cite{cavity,azimi_mousolou_magnon-magnon_2021,ding_magnon_2022}.
        
        The lifetime and entanglement challenges could find a common solution in applications based on an inverted magnetic state.
        The inverted magnetic state is a ferromagnet that is aligned antiparallel to a strong enough applied magnetic field, which overcomes the anisotropy.
        Therefore, it is in a state of maximum magnetic energy.
        The magnetic excitations on top of the inverted magnetic state are called antimagnons \cite{harms_antimagnonics_2024}.
        An antimagnon causes the magnetic moments to align more with the external magnetic field, therefore, lowering the total energy.
        Thus, an antimagnon has negative energy.
        This peculiarity causes an amplification of ordinary magnons when they scatter with antimagnons \cite{harms_enhanced_2022,wang_supermirrors_2024}, and entangled magnon-antimagnon pairs can be spontaneously excited \cite{bassant_entangled_2024,kleinherbers_entangling_2024}.
        These antimagnons are also involved in the Schwinger mechanism for magnons \cite{adorno_schwinger_2024} and topological states \cite{gunnink_2023,liu_tunable_2024} that allow for robust information processing.
        Therefore, they are a promising road to control and manipulate magnons to improve (quantum) magnonic devices.

        A ferromagnet that is in the inverted magnetic state is unstable and naturally relaxes to the ground state.
        Thus, to realize antimagnons, one has to stabilize the inverted magnetic state.
        This has been achieved before using spin transfer torque, as demonstrated in Ref. \cite{MagnetizationReversal}.
        More recently, both Ref. \cite{Kurebayashi_Barker_Yamazaki_Kushwaha_Stenning_Youel_Hou_Dion_Prestwood_Bauer_et_al._2026} and Ref. \cite{Karadza_Wang_Kercher_Noel_Legrand_Schlitz_Gambardella_2026} achieved the inverted magnetic state in a thin film through spin transfer torque and spin-orbit torque, respectively. 
        Here, we theoretically consider spin-orbit torque, which is achieved by considering an additional layer of heavy metal on top of the ferromagnet.
        This torque counteracts the Gilbert damping and dynamically stabilizes the ferromagnet opposite to the external magnetic field.
        
        In order to use antimagnons, it is necessary to understand their statistics.
        Since the inverted magnetic state is far out of equilibrium, unconventional fluctuations may arise.
        Typical fluctuations in a magnetic system arise from electron-magnon and electron-phonon scattering.
        In this setup, spin-orbit torque may cause additional fluctuations from Joule heating and shot noise in the spin current \cite{tserkovnyak2001shot,chudnovskiy2008spin}.
        However, spin-orbit torque competes with the Gilbert damping, and for different levels of competition, the magnetization can be more or less susceptible to these fluctuations.
        When the spin-orbit torque exactly opposes the Gilbert damping, the magnetic moment is very susceptible to fluctuations.
        Therefore, it could potentially be used as a stochastic bit, see Ref \cite{SB:debashis2020correlated,SB:debashis2022spin}.
        In this article, we compute the fluctuation classically and show that they are related to the magnetoresistance of the current in the heavy metal layer, which can be experimentally measured.
        In the low temperature limit, we identify a difference between fluctuations in the ground state and the inverted magnetic state, due to quantum effects.
        Here, we propose that a qubit can be used to measure the quantum fluctuations we predict here.

        The remainder of this paper is structured as follows.
        First, we introduce the setup in Sec. \ref{sec:Set-up}.
        In Sec. \ref{sec:classical}, we investigate the classical dynamics, which allows us to compute the influence of spin accumulation, and interfacial and bulk fluctuations on the fluctuations of the inverted magnetic state.
        Thereafter, in Sec. \ref{Sec:quantum}, the quantum description is taken into account to show the fundamental difference between antimagnon and magnon distribution and how it can be verified using a qubit.
        Finally, we conclude with a discussion in Sec. \ref{sec:Discussion and conclusions}.
    \section{Set-up: Inverted magnetic state}\label{sec:Set-up}
    \begin{figure}
        \centering
	\includegraphics[width = 0.6\columnwidth]{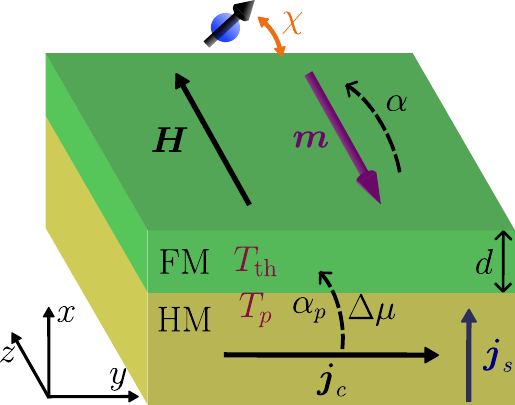}
	\caption{Set-up: A ferromagnet (FM) of thickness $d$ is coupled to a heavy metal (HM). An applied charge current $\boldsymbol{j}_c$ generates a spin current $\boldsymbol{j}_s$ in the HM, causing an interfacial spin accumulation $\Delta\mu$. Angular momentum is transferred into the FM with an efficiency governed by $\alpha_p$, inverting the magnetization $\boldsymbol{m}$ against its equilibrium position that is parallel to the applied magnetic field $\boldsymbol{H}$. Gilbert damping $\alpha$ works towards bringing $\boldsymbol{m}$ into its ground state. Temperatures present here are $T_p$ at the HM$|$FM interface and $T_\mathrm{th}$ in the FM bulk. A spin qubit is coupled with interaction strength $\chi$ for dispersive readout. \label{fig:Fig1}}
    \end{figure}
    We consider a ferromagnet (FM) of thickness $d$ with an applied magnetic field on top of a heavy metal (HM) (Fig.~\ref{fig:Fig1}). The energy of the FM with spin exchange coupling, easy-axis anisotropy and interaction with an applied magnetic field is given by~\cite{akhiezer1968, kittel1987} 
    \begin{equation}
		E_\mathrm{FM} = -J\sum_{\langle i,j\rangle} \boldsymbol{S}_i\cdot\boldsymbol{S}_j -K\sum_i\left(S_i^z\right)^2+ H\sum_i S_i^z, \label{eq:H_FM_spin}
    \end{equation}
    where the sum $\langle i,j\rangle$ denotes nearest neighbors, $\boldsymbol{S}_i$ the spin at lattice site $i$, $J$ the exchange interaction strength, $K$ the anisotropy energy, and $H>0$ an applied magnetic field along the $z$-direction in units of energy. The gyromagnetic ratio $\gamma<0$ is assumed to be negative here such that macrospin $\boldsymbol{S} = \sum_i\boldsymbol{S}_i$ and magnetization $\boldsymbol{m}$ anti-align. In the energetically stable ground state, the magnetization of the FM aligns with the external magnetic field $\boldsymbol{m}\parallel\boldsymbol{H}$.
    
    The FM is adjacent to a HM, such as platinum, that is a spin-Hall active material (Fig.~\ref{fig:Fig1}). The spin-Hall effect (SHE) is a phenomenon that stems from spin-orbit coupling (SOC), and describes the effect that an applied charge current $\boldsymbol{j}_c$ induces a spin current $\boldsymbol{j}_s$ perpendicular to $\boldsymbol{j}_c$~\cite{jungwirth2012}. The spin current $\boldsymbol{j}_s$ induces spin-flip scattering at the interface, such that spin-orbit torque (SOT) is exerted on the FM~\cite{duine2007}. The SOT excites the FM by effectively deviating the magnetization from its equilibrium position. For a suitable direction of the charge current, if the charge current is strong enough (larger than a critical current $|\boldsymbol{j}_\mathrm{crit}|$), the SOT brings the FM into a state of maximum magnetic energy. In this state the magnetization is antiparallel to the applied magnetic field such that the magnetization direction is inverted~\cite{harms_antimagnonics_2024}. This is an energetically unstable state, but the SOT compensates intrinsic damping that would bring the magnetization into the ground state. The inverted state is therefore dynamically stabilized by SOT. Additionally, we assume a temperature $T_\mathrm{th}$ in the FM bulk and a temperature $T_p$ at the HM$|$FM interface. These two temperatures may be different and are related to bulk and interfacial thermal fluctuations. In the following, we analyze the fluctuations of the inverted magnetic state within a classical framework treating the dynamics with stochastic Landau-Lifshitz-Gilbert equations in Sec.~\ref{sec:classical}. We then turn our attention to a low temperature and single domain set-up and use a master equation approach to understand the quantum fluctuations of the inverted magnetic state in Sec.~\ref{Sec:quantum}.
	
    \section{Classical dynamics\label{sec:classical}}
    In this section, we analyze the fluctuations of the inverted magnetic state within a classical framework, assuming the set-up as in Fig.~\ref{fig:Fig1}. Our goal is to understand the influence of the adjacent HM on the fluctuations of the FM. While the HM is responsible for stabilizing the energetically unstable inverted state, we are also interested in the fluctuations due to SOT and their influence in relation to the system size. 
    
    \subsection{Stochastic Landau-Lifshitz-Gilbert equation\label{sec:LLG}}

    The dynamics of the magnetization $\boldsymbol{m}$ and hence the macrospin $\boldsymbol{S}$, assuming $\boldsymbol{S}_i$ uniform, are described via the Landau-Lifshitz-Gilbert (LLG) equation~\cite{stancil2009}. Using the spin density unit vector $\boldsymbol{n}$, with the spin density $s$, the LLG is formulated as ($\hbar =1$)
    \begin{equation}
        \dot{\boldsymbol{n}} = \boldsymbol{n}\times \frac{\delta E_\mathrm{FM}}{\delta \boldsymbol{n}} - \alpha \boldsymbol{n} \times \dot{\boldsymbol{n}}, \label{eq:LLG}
    \end{equation}
    with Gilbert damping $\alpha$. Here, we consider the classical limit, such that the energy functional $E_\mathrm{FM}$ from equation Eq.~\eqref{eq:H_FM_spin} has to be considered in the continuum limit in Eq.~\eqref{eq:LLG}. This can be achieved by expansion of spin interactions~\cite{ stancil2009}. Here, the energy functional is given by $E_\mathrm{FM} = \int\mathrm{d}V\left\{A(\nabla\boldsymbol{n})^2 - Hn_z -Kn_z^2/2\right\}$. In order to capture the thermal fluctuations of the system, we introduce a stochastic field $\boldsymbol{h}$ and use the stochastic LLG equation. Using $E_\mathrm{FM}$ [Eq.~\eqref{eq:H_FM_spin}] in the continuum limit, the stochastic LLG explicitly reads~\cite{zheng2017}
    \begin{align}
\left(1+\alpha\boldsymbol{n}\times\right)\dot{\boldsymbol{n}}+\boldsymbol{n}\times\left(\boldsymbol{H}+\boldsymbol{h}-Kn_z\boldsymbol{z}\right)-A\boldsymbol{n}\times\nabla^{2}\boldsymbol{n} & =0, \label{eq:sLLG}
    \end{align}
    with the applied magnetic field $\boldsymbol{H}=H\boldsymbol{z}$, spin stiffness $A=Ja^2$ that is related to the exchange integral $J$ and lattice constant $a$. The easy-axis anisotropy $K$ in linear spin wave theory takes on a form that is equivalent to a magnetic field, hence its effects can be taken into account by defining an effective magnetic field $ \boldsymbol{H}_\text{eff} = \boldsymbol{H} - Kn_z\boldsymbol{e}_z$. Furthermore, hard axis anisotropy is considered in Ref.~\cite{harms_antimagnonics_2024}. Therein we see that hard axis anisotropy causes ellipticity.
    Ellipticity is not considered in this article, however, our formulation can be extended to take this into account by using a Bogoliubov transformation as in Ref.~\cite{harms_antimagnonics_2024}.

    An applied charge current $\boldsymbol{j}_c$ along the $y$-direction on the HM causes a spin current $\boldsymbol{j}_s$ along the $x$-direction where the HM interfaces with the FM. The spin current in the FM is $j_s = -As\boldsymbol{n}\times\partial_{x}\boldsymbol{n}$ and induces a spin accumulation $\Delta\boldsymbol{\mu}=-\Delta\mu\boldsymbol{z}$ along the HM$|$FM interface with $\Delta\mu = \mu_\uparrow - \mu_\downarrow$~\cite{bender2017, hoffman2013}. We therefore impose the boundary condition at the HM$|$FM interface ($x=0$)
	\begin{align}
		j_{s}\!&\left(x=0\right)= \nonumber\\ &-\left[\frac{g^{\uparrow\downarrow}}{4\pi}\left(\boldsymbol{n}\times\left(\boldsymbol{n}\times\Delta\boldsymbol{\mu}\right)+\boldsymbol{n}\times\dot{\boldsymbol{n}}\right)+\boldsymbol{n}\times\boldsymbol{h}_{L}^{\prime}\right]_{x=0}, \label{eq:boundary0}
	\end{align}
    where $\boldsymbol{h}_{L}$ denotes the interfacial stochastic field that accounts for the fluctuations in the spin current and $g^{\uparrow\downarrow}$ the spin mixing conductance. At the interface with the vacuum ($x=d$), the spin current vanishes such that we impose the condition $j_{s}\!\left(x=d\right)= 0$. 

    Here, we want to analyze small deviation from the equilibrium orientation ${n}$. In the inverted magnetic state, we linearise around $\boldsymbol{n}=+\boldsymbol{z}$. Defining the complex field $\psi\left(\boldsymbol{x},t\right)=n\left(\boldsymbol{x},t\right)\sqrt{s/2}$ with $n=n_{x}+in_{y}$, we can conveniently quantify the deviations from the equilibrium, such that the linearization of the stochastic LLG [Eq.~\eqref{eq:sLLG}] becomes
	\begin{align}
            A\left(\nabla^{2}+\frac{\omega_0+i\left(1+i\alpha\right)\partial_{t}}{A}\right)\psi\!\left(\boldsymbol{x},t\right) & =\sqrt{\frac{s}{2}}h\!\left(\boldsymbol{x},t\right), \label{eq:sLLG_psi}
	\end{align}
    with the complex stochastic bulk field $h=h_{x}+ih_{y}$ and $\omega_0 = H - K$. Correspondingly, the boundary condition at $x=0$ [Eq.~\eqref{eq:boundary0}] can be linearized around the equilibrium, such that
	\begin{align} A\partial_{x}\psi\!\left(\boldsymbol{x},t\right)|_{x=0}+i\frac{g^{\uparrow\downarrow}\left(i\partial_{t}-\Delta\mu\right)}{4\pi s}&\psi\!\left(\boldsymbol{x},t\right)\rvert_{x=0}\!\!\!\!
		&=\frac{h_{L}\!\left(\boldsymbol{\rho},t\right)}{\sqrt{2s}},\label{eq:cond1}
	\end{align}
    where we defined the two-dimensional coordinate $\boldsymbol{\rho}$ that is related to the three-dimensional coordinate $\boldsymbol{x}$ via $\boldsymbol{x}=\left(x,\boldsymbol{\rho}\right)$. The boundary condition at $x=d$ in terms of the complex field $\psi\left(\boldsymbol{x},t\right)$ becomes $\partial_{x}\psi\left(\boldsymbol{x},t\right)\rvert_{x=d}=0$. Additionally, the bulk stochastic field $\boldsymbol{h}$ and interfacial stochastic field $\boldsymbol{h}_L$ fulfill the fluctuation-dissipation theorem (FDT)~\cite{zheng2017}
	\begin{align}
		\left\langle h^{*}\!\left(x,\boldsymbol{q},\omega\right)h\!\left(x^{\prime},\boldsymbol{q}^{\prime},\omega^{\prime}\right)\right\rangle =~~&  \nonumber \\ \frac{2}{s}\frac{\left(2\pi\right)^{3}\alpha\omega}{\tanh(\frac{\omega}{2T_\mathrm{th}})}\delta\!\left(x-x^{\prime}\right)&\delta\!\left(\boldsymbol{q}-\boldsymbol{q}^{\prime}\right)\delta\!\left(\omega-\omega^{\prime}\right),\label{eq:FDT}\\
		\left\langle h_{L}^{*}\!\left(\boldsymbol{q},\omega\right)h_{L}\!\left(\boldsymbol{q}^{\prime},\omega^{\prime}\right)\right\rangle  =~~~~~~& \nonumber \\
		\frac{2\left(2\pi\right)^{3}\frac{g^{\uparrow\downarrow}}{4\pi}\left(\omega-\Delta\mu\right)}{\tanh(\frac{\omega-\Delta\mu}{2T_p})} &\delta\!\left(\boldsymbol{q}-\boldsymbol{q}^{\prime}\right)\delta\!\left(\omega-\omega^{\prime}\right), \label{eq:FDT_interface}
	\end{align}
   with the temperature of the bulk FM (HM interface) $T_{\mathrm{th}(p)}$ (see Fig.~\ref{fig:Fig1}). We proceed by solving for the complex field of deviations $\psi\left(\boldsymbol{x},t\right)$. Our differential equation [Eq.~\eqref{eq:sLLG_psi}] with boundary condition [Eq.~\eqref{eq:cond1}] can be formulated in one equation of the form 
	\begin{align}
		\mathcal{L}\psi\!\left(x,\boldsymbol{\rho},t\right) & =h_{\text{tot}}\!\left(x,\boldsymbol{\rho},t\right),
	\end{align}
	with the differential operator 
	\begin{align}
		\!\!\mathcal{L} = A\nabla^{2}+\omega_0+i\left(1+i\alpha\right)\partial_{t}+i\frac{g^{\uparrow\downarrow}\left(i\partial_{t}-\Delta\mu\right)}{4\pi s}\delta\!\left(x\right), 
	\end{align}
	and the total fluctuations given by $h_{\text{tot}}\!\left(x,\boldsymbol{\rho},t\right)=\sqrt{s/2}h\!\left(x,\boldsymbol{\rho},t\right)+\delta\!\left(x\right)h_{L}\!\left(\boldsymbol{\rho},t\right)/\sqrt{2s}$. We solve this equation using the Green's function formalism.
    The Green's function satisfies $\mathcal{L}G\!\left(x,x^{\prime},\boldsymbol{\rho}-\boldsymbol{\rho}^{\prime},t-t^{\prime}\right)=\delta\!\left(x-x^{\prime}\right)\delta\!\left(\boldsymbol{\rho}-\boldsymbol{\rho}^{\prime}\right)\delta\!\left(t-t^{\prime}\right)$, then the solution can be expressed in terms of the stochastic field and the Green's function, $\psi\!\left(x,\boldsymbol{\rho},t\right)=\int_{0}^{d}\text{d}x^{\prime}\int\text{d}t^{\prime}\int\text{d}^{2}\boldsymbol{\rho}^{\prime}G\!\left(x,x^{\prime},\boldsymbol{\rho}-\boldsymbol{\rho}^{\prime},t-t^{\prime}\right)h_{\text{tot}}\!\left(x^{\prime},\boldsymbol{\rho}^{\prime},t^{\prime}\right)$. Performing the Fourier transform for the $y$ -and $z$-components, our ansatz becomes~\cite{zheng2017}
	\begin{align}
		\left(A\partial_{x}^{2}-A\kappa^{2}+i\frac{g^{\uparrow\downarrow}\left(\omega-\Delta\mu\right)}{4\pi s}\delta\!\left(x\right)\right)&G\!\left(x,x^{\prime},\boldsymbol{q},\omega\right)\nonumber\\& =\delta\!\left(x-x^{\prime}\right),\label{eq:DGL}
	\end{align}
	with 
	\begin{align}
		\kappa^{2} & =\boldsymbol{q}^{2}-\frac{\omega_0+\omega+i\alpha\omega}{A}, \label{eq:kappa}
	\end{align}
	and the Green's function Fourier transform $G\!\left(x,x^{\prime},\boldsymbol{q},\omega\right)$. Following the formalism from~\cite{zheng2017}, we find that the solution for Eq.~\eqref{eq:DGL} is given by
	\begin{align}
		G\!\left(x,x^{\prime},\boldsymbol{q},\omega\right)  =\frac{1}{AW}[&\chi_{L}\!\left(x^{\prime}\right)\chi_{R}\!\left(x\right)\Theta\!\left(x-x^{\prime}\right)\nonumber \\
		&+\chi_{L}\!\left(x\right)\chi_{R}\!\left(x^{\prime}\right)\Theta\!\left(x^{\prime}-x\right)], \label{eq:Green_sol}
	\end{align}
	with the functions 
	\begin{align}
		\chi_{L}\!\left(x\right) & =-i\frac{g^{\uparrow\downarrow}\left(\omega-\Delta\mu\right)}{4\pi s}\sinh(\kappa x)+A\kappa\cosh(\kappa x),\\
		\chi_{R}\!\left(x\right) & =\cosh(\kappa x-\kappa d),
	\end{align}
	and the Wronskian 
	\begin{align}
		W & =\kappa\left[i\frac{g^{\uparrow\downarrow}}{4\pi s}\left(\omega-\Delta\mu\right)\cosh(\kappa d)-A\kappa\sinh(\kappa d)\right].\label{eq:wronskian}
	\end{align}
	Having determined the Green's function $G\!\left(x,x^\prime, \boldsymbol{q},\omega, \right)$ [Eq.~\eqref{eq:Green_sol}], we can now compute $\psi\!\left(\boldsymbol{x},t\right)$ by integrating the product of $G\!\left(\boldsymbol{x},\boldsymbol{x}^\prime, t-t^\prime\right)$ and $h_\mathrm{tot}\!\left(\boldsymbol{x},t\right)$. Using the correlators from the FDT [Eqs.~\eqref{eq:FDT} and \eqref{eq:FDT_interface}], we obtain an explicit, but lengthy, expression for the average magnon density $\left\langle \psi^{*}\!\left(x,\boldsymbol{\rho},t\right)\psi\!\left(x,\boldsymbol{\rho},t\right)\right\rangle $, which can be found in Appendix~\ref{sec:appendix_integrand}. In the thin film limit, $\kappa d\ll1$, the expression for the average magnon density $\left\langle \psi^{*}\!\left(x,\boldsymbol{\rho},t\right)\psi\!\left(x,\boldsymbol{\rho},t\right)\right\rangle $ simplifies to $\left\langle \psi^{*}\!\left(x,\boldsymbol{\rho},t\right)\psi\!\left(x,\boldsymbol{\rho},t\right)\right\rangle=
	\int_{-\infty}^{\infty}\frac{\text{d}\omega}{2\pi}\int\frac{\text{d}^{2}\boldsymbol{q}}{\left(2\pi\right)^{2}}I\!(\omega,\boldsymbol{q})$, where the integrand $I\!(\omega,\boldsymbol{q})$ is given by $I\!(\omega,\boldsymbol{q}) = I_p\!(\omega,\boldsymbol{q})+I_{\mathrm{th}}\!(\omega,\boldsymbol{q})$ with the summands
	\begin{align}
		I_p\!(\omega,\boldsymbol{q}) =& \frac{\frac{2}{d}\alpha_p\left(\omega-\Delta\mu\right)\left(n_{B}^p\!\left(\omega-\Delta\mu\right)+\frac{1}{2}\right)}{\left(\omega_0+\omega-A\boldsymbol{q}^{2}\right)^{2}+\left(\alpha\omega+\alpha_p\left(\omega-\Delta\mu\right)\right)^{2}}, \label{eq:integrand1} \\ 
		I_{\mathrm{th}}\!(\omega,\boldsymbol{q}) = & \frac{\frac{2}{d}\alpha\omega\left(n_{B}^\mathrm{th}\!\left(\omega\right)+\frac{1}{2}\right)}{\left(\omega_0+\omega-A\boldsymbol{q}^{2}\right)^{2}+\left(\alpha\omega+\alpha_p\left(\omega-\Delta\mu\right)\right)^{2}}, \label{eq:integrand2}
	\end{align}
	with $\alpha_p d=g^{\uparrow\downarrow}/4\pi s$ and the Bose-Einstein distributions $n_{B}^{p/\mathrm{th}}\!\left(\epsilon\right) = \left[\exp(\epsilon\beta_{p/\mathrm{th}})-1\right]^{-1}$ with the inverse temperatures $\beta_{p(\mathrm{th})}=1/k_BT_{p(\mathrm{th})}$. The $\omega$-integral of $I\!(\omega,\boldsymbol{q})$ [Eqs.~\eqref{eq:integrand1} and \eqref{eq:integrand2}] is performed analytically via contour-integration,
    giving rise to a stability condition for the spin accumulation $\Delta\mu/\omega_0 < - (\alpha + \alpha_p)/\alpha_p$~\cite{harms_antimagnonics_2024}\footnote{The analytical integration reveals and highlights the importance of pumping SOT for the stability of the state: Setting $\alpha_p =0$ results in negative magnon density $\left\langle \psi^{*}\!\left(\boldsymbol{x},t\right)\psi\!\left(\boldsymbol{x},t\right)\right\rangle$, hence an unstable state.}. The explicit expression can be found in Appendix~\ref{sec:appendix_integrand}. This instability captures the competition between Gilbert damping from the bulk ($\alpha$) and spin injection through the interface, governed by $\alpha_p$. If the stability condition is fulfilled, the pumping compensates the damping, and the inverted state is dynamically stabilized. Note that this stability condition has been derived before using a homogeneous Green's function approach in Ref.~\cite{harms_antimagnonics_2024}.
	
	The expressions in Eqs.~\eqref{eq:integrand1} and Eqs.~\eqref{eq:integrand2} allow us to compute the spin Hall magnetoresistance in the HM. The resistivity that is longitudinal to the applied charge current $\boldsymbol{j}_c$ (see Fig.~\ref{fig:Fig1}) is approximately given by~\cite{chen2013}
	\begin{align}
		\rho_\text{long} & \approx\rho+\Delta\rho_{0}+\Delta\rho_{1}\left(1-n_{z}^{2}\right), \label{eq:resistance}
	\end{align}
	where $\rho$ denotes the resistivity of the bare HM, $\Delta\rho_0$ quantifies a material dependent change in resistivity due to spin-Hall magnetoresistance, and $\Delta\rho_1$ quantifies the state dependent part of the magnetoresistance that depends on $n_{z}^{2}$ at the interface $x=0$. Due to the relation
	\begin{align}
		1-n_{z}^{2} &  \approx\frac{2}{s}\left\langle \psi^{*}\psi|_{x=0}\right\rangle, \label{eq:res_fluct}
    \end{align}
    We argue that the magnon or antimagnon density is directly related to the resistivity, hence can be measured in transport experiments.
    \subsection{Results \label{sec:classical_results}}
    \begin{figure}
	\centering
        \includegraphics[width=0.99\linewidth]{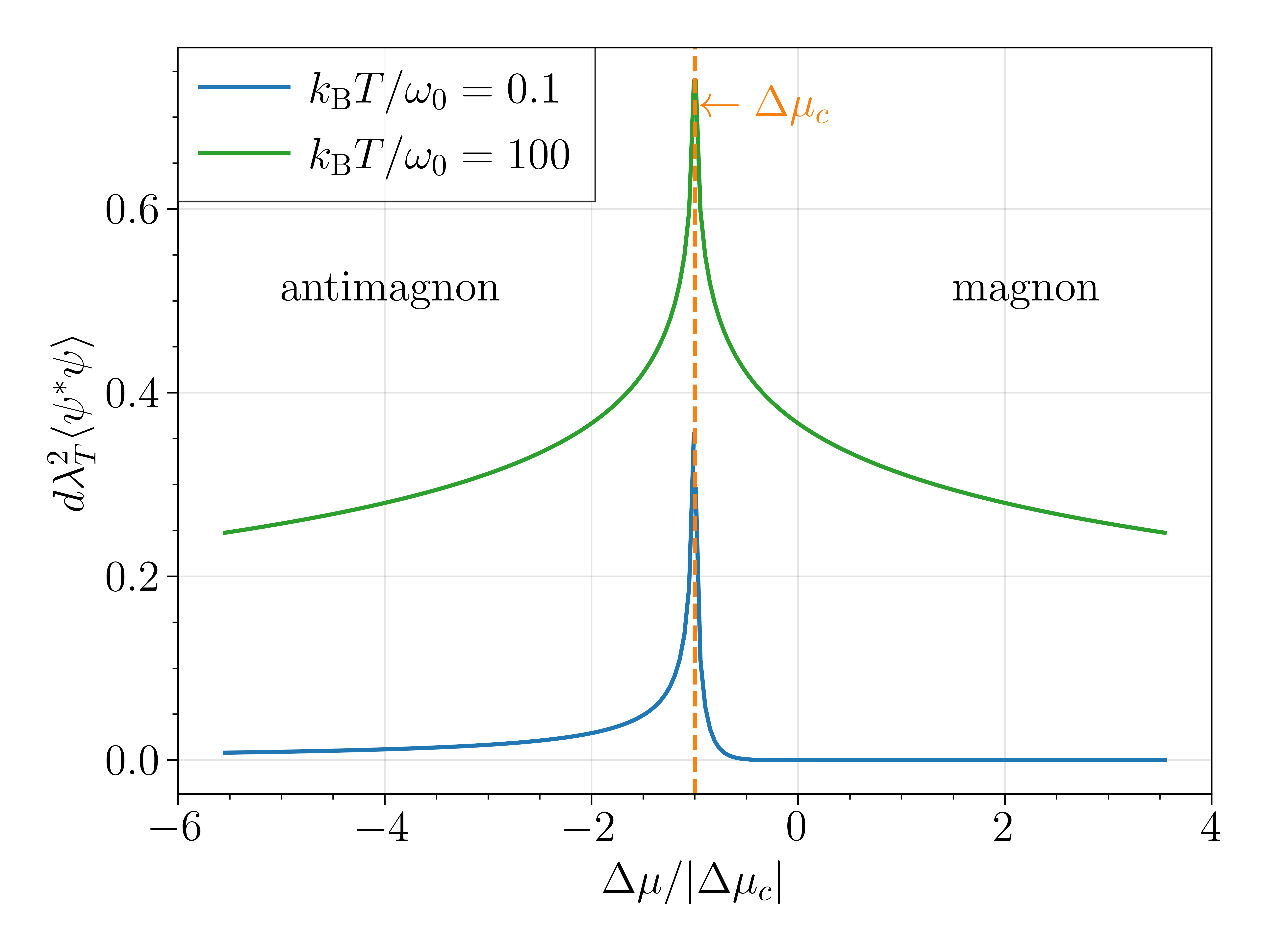}
	\caption{Occupation $d\lambda_T^2\left\langle\psi^*\psi\right\rangle$ is plotted as a function of spin accumulation  $\Delta\mu$ for temperatures $T/\omega_0 = 0.1$ (blue) and $T/\omega_0 = 100$ (green) with $T_p=T_{\mathrm{th}}=T$ and $\alpha=0.001$, $\alpha_p=0.01$ and $d/\xi = 0.09$ fixed. The orange dashed line marks the critical spin accumulation $\Delta\mu_c/\omega_0 = -(\alpha+\alpha_p)/\alpha_p$. In the regime $\Delta\mu<\Delta\mu_c$ ($\Delta\mu>\Delta\mu_c$), we linearize around the inverted state (ground state) such that excitations are of antimagnonic (magnonic) nature.}
		\label{fig:Fig2}
	\end{figure}
	In the following, we study the magnon density $\left\langle \psi^{*}\!\left(x,\boldsymbol{\rho},t\right)\psi\!\left(x,\boldsymbol{\rho},t\right)\right\rangle $ by performing the $\omega$ and $\boldsymbol{q}$-integration numerically and multiply by the volume $d\lambda_T^2$ where $\lambda_T$ denotes the magnon thermal wave length $\lambda_T = \sqrt{A/T}$. Our linearization of the LLG is valid if $d\lambda_T^2 \left\langle \psi^{*}\!\left(x,\boldsymbol{\rho},t\right)\psi\!\left(x,\boldsymbol{\rho},t\right)\right\rangle <1$ and $d/\xi \ll 1$ with $\xi = \sqrt{A/\omega_0}$. The latter condition ensures that our assumption of a homogeneous inverted state is valid since our linearization cannot capture domain wall formation. Since the relevant physical observable $\rho_{\mathrm{long}}$ [Eq.~\eqref{eq:resistance}] is related to $\left\langle \psi^{*}\psi|_{x=0}\right\rangle$ [Eq.~\eqref{eq:res_fluct}], we will evaluate the average magnon density at the interface $x=0$ and abbreviate it by $\left\langle \psi^{*}\psi\right\rangle$. Also, the FDT of our stochastic field from Eqs.~\eqref{eq:FDT} and \eqref{eq:FDT_interface} are proportional to $n_{B}^{p/\mathrm{th}}\!\left(\epsilon\right)+1/2$. The half is due to the classical treatment of vacuum fluctuations. Quantum mechanically, we employ normal ordering of the operators, which causes the half to disappear, see Ref. \cite{stoof1999coherent}. We have taken the quantum mechanical result to ensure that our integrals converge.
    
	In Fig.~\ref{fig:Fig2}, we plot the dependence of the normalized magnon density $d\lambda_T^2 \left\langle \psi^{*}\psi\right\rangle$ on spin accumulation $\Delta\mu$ for two different temperatures where the bulk FM and the interface have the same temperature $T_p=T_{\mathrm{th}} \equiv T$.  Fig.~\ref{fig:Fig2} shows two regimes: $\Delta\mu < \Delta\mu_c$ and $\Delta\mu > \Delta\mu_c$ with the critical value $\Delta\mu_c/\omega_0 = -(\alpha + \alpha_p)/\alpha_p$. The computation for $\Delta\mu < \Delta\mu_c$ has been performed linearizing around the inverted state whereas in the $\Delta\mu > \Delta\mu_c$ regime, we linearize around the true ground state.
    We see that both lower temperature and larger $|\Delta\mu-\Delta \mu_c|$ suppresses fluctuations.
    At the critical value $\Delta\mu_c$, the curves diverge in both regimes, giving rise to the instability. \alrcom{
    We find that $\langle \psi^*\psi\rangle \propto \ln(|\Delta\mu-\Delta\mu_c|)$ is a logarithmic divergence and} can be interpreted as a driven phase transition: Large spin accumulation $\Delta\mu$ suppresses the thermal fluctuations of the ground state. Lowering $\Delta\mu$ reduces the suppression of fluctuations while going to negative $\Delta\mu<0$ enhances the fluctuations until fluctuations diverge at $\Delta\mu_c$ and the state is destabilized. In the regime $\Delta\mu<\Delta\mu_c$, the spin accumulation stabilizes the inverted state, while smaller $\Delta\mu$ reduce the state's fluctuations again. The curve for $T/\omega_0=100$ is symmetric around the critical spin accumulation $\Delta\mu_c$, the curve for small temperature $T/\omega_0=0.1$ is not, suggesting that the inverted state exhibits more fluctuations than the ground state. We confirm this finding for small temperatures with quantum mechanical calculations in Sec.~\ref{Sec:quantum}.
	
\begin{figure}
            \centering
            \includegraphics[width=0.99\linewidth]{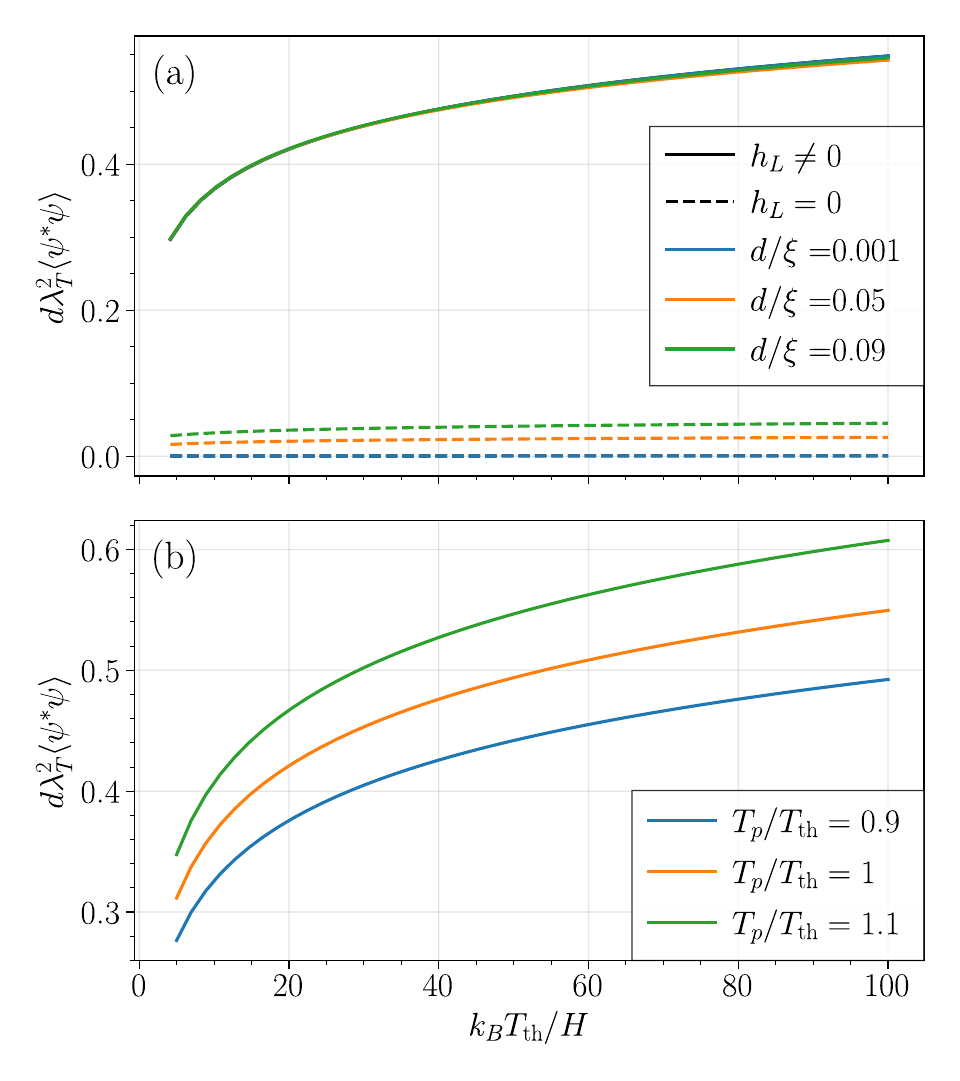}
            \caption{Occupation $d\lambda_T^2\left\langle\psi^*\psi\right\rangle$ is plotted as a function of bulk temperature $T_{\mathrm{th}}$. (a) We compare the full dynamics with $h_l\neq0$ (solid) with neglecting fluctuations due to SOT $h_L=0$ (dashed) for three values of FM thickness $d/\xi =0.001$ (blue), $d/\xi=0.05$ (orange) and $d/\xi=0.09$ (green), with $\alpha = \alpha_pd/\xi =0.00001$, $\Delta\mu/\omega_0 = -1.1(\alpha+\alpha_p)/\alpha_p$ and $T_p=T_\mathrm{th}$. (b) We compare the full dynamics of 3 different ratios of $T_p/T_\mathrm{th}$: $T_p/T_\mathrm{th}=0.9$ (blue), $T_p/T=1$ (orange) and $T_p/T_\mathrm{th}=1.1$ (green) for $d/\xi=0.095$, $\alpha=0.001$, $\alpha_pd/\xi=0.01$ and $\Delta\mu/\omega_0 = -1.1(\alpha+\alpha_p)/\alpha_p$.}
            
            \label{fig:Fig3}
        \end{figure}

	In Fig.~\ref{fig:Fig3}, we explore the influence of the fluctuations due to SOT, denoted by $h_L$, and interface temperature $T_p$. In Fig.~\ref{fig:Fig3}(a), we compare the normalized magnon density $d\lambda_T^2\left\langle \psi^{*}\psi\right\rangle$ taking into account the fluctuations due to SOT ($h_L\neq0$) with neglecting them ($h_L=0$) for three different thicknesses $d$. We perform this analysis in the regime where $\alpha = \alpha_p d/\xi = 0.00001$, so both the bulk and interfacial effects on the state are weak. From Fig.~\ref{fig:Fig3}(a), we deduce that the fluctuations due to SOT have a non-negligible effect on the magnon density, but this effect weakens with larger thicknesses $d$. The ratios $r = \left\langle \psi^{*}\psi\right\rangle|_{h_L=0}/\left\langle \psi^{*}\psi\right\rangle$, which expresses as the difference between solid and dashed lines in Fig.~\ref{fig:Fig3}(a), concretely read $r|_{d/\xi=0.001} = 0.001$, $r|_{d/\xi=0.05} = 0.048$ and $r|_{d/\xi=0.09} = 0.083$. From the analytic $\omega$-integration, we estimate the ratio in the large temperature limit when $T_p = T_\mathrm{th}$ and find that $r\approx 1/\Delta\mu_c = [1 + \alpha_p\xi/(\alpha d)]^{-1}$. In Fig.~\ref{fig:Fig3}(b), we plot occupation $d\lambda_T^2\left\langle \psi^{*}\psi\right\rangle$ as a function of bulk temperature $T_{\mathrm{th}}$ for three different ratios between interfacial and bulk temperature $T_p/T_{\mathrm{th}}$. We conclude that the interface can function as a cool reservoir suppressing fluctuations ($T_p/T_{\mathrm{th}}=0.9$) or as a hot reservoir enhancing fluctuations ($T_p/T_{\mathrm{th}}=1.1$). However, it should be noted here, that our model does not take into account an additional torque from the spin-Seebeck effect due to the interaction between thermal magnons and the ground state magnons~\cite{bender2014}.
    Therefore, we limit our analysis to ratios of $T_p/T_{\mathrm{th}}$ close to $1$.

	\section{Quantum description\label{Sec:quantum}}
	Motivated by the previous section, we will consider the low-temperature behavior in a quantum mechanical framework. For this purpose, we consider the FM Hamiltonian obtained from energy functional $E_\mathrm{FM}$ [Eq.~\eqref{eq:H_FM_spin}] by substituting the spin vectors by operators $\boldsymbol{S}_i\rightarrow\hat{\boldsymbol{S}}_i$. Here, we consider the inverted magnetic state and choose our quantization axis accordingly. Choosing the applied magnetic field along the positive
	$\hat{\boldsymbol{e}}_{z}$ axis, we use the following linearized Holstein-Primakoff transformation, hence quantization of spin operators~\cite{harms_antimagnonics_2024,holstein1940}
	\begin{align}
		\hat{S}_{i}^{z} & \approx S-\hat{\psi}_{i}^{\dagger}\hat{\psi}_{i},\label{eq:Sz}\\
		\hat{S}_{i}^{+} & \approx \sqrt{2S}\hat{\psi}_{i},\label{eq:S+}
	\end{align}
	where $\hat{\psi}_{i}^{\dagger}$ denotes the creation operator of a negative energy excitation on the inverted magnetic state at lattice site $i$ and $\langle\hat{\psi}_{i}^{\dagger}\hat{\psi}_{i}\rangle\ll 2S$. Note here that we choose the quantization axis for the spin operators along the magnetic field, since we consider negative gyromagnetic ratio $\gamma<0$. It has been shown explicitly in Ref.~\cite{harms_antimagnonics_2024} that with this quantization, the Hamiltonian derived from $E_\mathrm{FM}$ [Eq.~\eqref{eq:H_FM_spin}] becomes 
	\begin{equation}
		\hat{\mathcal{H}}_\mathrm{FM} = -\omega_0\hat{\psi}^\dagger \hat{\psi},\label{eq:H_quant}
	\end{equation}
    where $\omega_0 = H-K$ and $[\hat{\psi}, \hat{\psi}^\dagger]=1$. Note that we only took the uniform mode into account with the creation operator $\hat{\psi}^\dagger$. Because of the negative energies, the Lindblad operators of the Hamiltonian $\hat{\mathcal{H}}_\mathrm{FM} $ [Eq.~\eqref{eq:H_quant}] are $A\!\left(\omega_0\right)=\hat{\psi}^\dagger$ and $A\!\left(-\omega_0\right)=\hat{\psi}$. Here we can already note that the creation and annihilation operators changed their typical role for bosons~\cite{breuer2007}. 
	
	The system is under the influence of SOT and Gilbert damping. While SOT stabilizes the inverted state, Gilbert damping destabilizes it and works towards bringing the system into the energetically favourable FM ground state. We model these two influences with thermal baths of harmonic oscillators. In the following, we look at the two baths and their effects separately and then bring them together in the end.   
		
	We start with the bulk lattice bath, resulting in Gilbert damping. We model it as a bath of harmonic oscillators with the bath Hamiltonian $\hat{\mathcal{H}}_\mathrm{th} = \sum_i \omega_i \hat{a}^\dagger_i\hat{a}$ with bath operators $\hat{a}_i$. The bath density operator is given by $\rho_\mathrm{th} = \exp(-\beta_\mathrm{th}\hat{\mathcal{H}}_\mathrm{th})/Z_\mathrm{th}$ with the 
	partition function $Z_\mathrm{th}$. Following the master equation approach for a thermal bath, we arrive at the master equation
	\begin{align}
		\frac{\text{d}}{\text{d}t}p_{n} =&\gamma_{\text{th}}\left(n_{\text{th}}+1\right)\left(np_{n-1}-\left(n+1\right)p_{n}\right)\nonumber\\&+\gamma_{\text{th}}n_{\text{th}}\left(\left(n+1\right)p_{n+1}-np_{n}\right), \label{eq:bulk_bath}
	\end{align}
	with $\gamma_{\text{th}} =\alpha\omega_0$ and $n_{\text{th}} =n^\mathrm{th}_{\text{B}}\left(\omega_0\right)$. Note here that our master equation [Eq.~\eqref{eq:bulk_bath}] differs from the usual master equation for bosons in contact with a thermal bath since our Lindblad operators $A\!\left(\omega_0\right)$ are unconventional~\cite{breuer2007, walls2008}. In particular, Eq.~\eqref{eq:bulk_bath} demonstrates that emission and absorption switch their role compared to the typical thermal equilibrium, such that our system has spontaneous absorption.  We find that the stationary solution of Eq.~\eqref{eq:bulk_bath} is given by $p_{n} =p_{0}\left[\left(n_{\text{th}}+1\right)/n_{\text{th}}\right]^{n}$ which is divergent and hence leads to an instability. This is expected since the inverted state is destabilized by Gilbert damping that favours the energetically stable FM ground state.
	
    The SOT has to be handled with care in this formalism. We start from a Hamiltonian describing the HM as a sea of electrons $\hat{\mathcal{H}}_e = \sum_{\boldsymbol{k},\sigma = \uparrow\downarrow} \epsilon_{\boldsymbol{k}} \hat{c}_{\boldsymbol{k},\sigma}^\dagger\hat{c}_{\boldsymbol{k},\sigma}$ where $\hat{c}_{\boldsymbol{k},\sigma}$ represents the annihilation operator of an electron with momentum $\boldsymbol{k}$ and spin $\sigma$. Identifying a spin operator $\hat{\boldsymbol{s}}$ with $\hat{s}^z_i = \frac{1}{2}(\hat{c}_{i\uparrow}^\dagger\hat{c}_{i\uparrow} - \hat{c}_{i\downarrow}^\dagger\hat{c}_{i\downarrow})$ and $\hat{s}^+_i = \frac{1}{2}\hat{c}_{i\uparrow}^\dagger\hat{c}_{i\downarrow}$,with $\hat{s}^\pm_i = \hat{s}^x_i \pm i\hat{s}^y_i$ and $i$ denoting the lattice site, allows us to express the scattering process between the electrons of the HM and the FM as a spin exchange interaction $\hat{\mathcal{H}}_{\mathrm{ex}} = -J_{\mathrm{e-m}}\sum_{i} \hat{\boldsymbol{S}}_i \cdot \hat{\boldsymbol{s}}_i$~\cite{wang2015a, dusabirane2024}.
    The process that exerts angular momentum on the FM is $\propto \hat{c}_{i\downarrow}^\dagger \hat{c}_{i\uparrow}$, noting that this process dominates if the spin current $\boldsymbol{j}_s$ is polarized accordingly. This spin-flip scattering process can be interpreted as the creation of an electron-hole pair and hence re-expressed in terms of bosons. Choosing the same quantization axis as in Eqs.~\eqref{eq:Sz} and \eqref{eq:S+}, we can interpret the HM as a thermal bath of inverted harmonic oscillators $\hat{\mathcal{H}}_p=-\sum_j \omega_j \hat{b}_j^\dagger \hat{b}_j$ with bath operators $\hat{b}_j$. 
    The negative energies here have their origin in the energy difference between the incoming and outgoing electron. The spin polarization of the incoming electron is not favored by the applied magnetic field and hence has a higher energy than the outgoing electron. The energy difference gets transferred into the FM, which is a result driven to a state of maximum magnetic energy. 
    Fluctuations in the scattering process, e.g. the lack of scattering, lowers the energy and creates an antimagnon. The spin accumulation $\Delta\mu$ is treated as a chemical potential~\cite{harms_antimagnonics_2024} in the grand canonical ensemble $\rho_p = \exp(-\beta_p\left[{\hat{\mathcal{H}_p} - \Delta\mu\hat{\mathcal{N}}}\right])/Z_p$ with $\hat{\mathcal{N}}=\sum_j \hat{b}_j^\dagger \hat{b}_j$ and the partition function $Z_p =\prod_{j}\left[1-\exp\left(\beta_p\left(\omega_{j}+\Delta\mu\right)\right)\right]^{-1}$~\cite{landau1969}. 
    Then we evaluate the correlators of the bath oscillators as~\cite{walls2008}
	\begin{align}
		\left\langle \hat{b}^{\dagger}(\omega_j)\hat{b}(\omega_{j^\prime}) \right\rangle _{p} & =n_{\mathrm{B}}^p\!\left(-\left(\omega_{j}+\Delta\mu\right)\right)\delta\!(\omega_j-\omega_{j^\prime}).
	\end{align}
    Using this relation, we find that the master equation with SOT is given by 
    \begin{align}
    	\frac{\text{d}}{\text{d}t}p_{n} =&\gamma_{p}n_{p}\left(np_{n-1}-\left(n+1\right)p_{n}\right)\nonumber\\&+\gamma\left(n_{p}+1\right)\left(\left(n+1\right)p_{n+1}-np_{n}\right),\label{eq:SOT_bath}
    \end{align}
	with $\gamma_{p} =-\alpha_{p}\left(\omega_0+\Delta\mu\right)$ and $n_{p} =n^p_{\text{B}}\!\left(-\left(\omega_0+\Delta\mu\right)\right)$. The solution of Eq.~\eqref{eq:SOT_bath} is given by $p_{n} =p_{0}\left(\frac{n_{p}}{n_{p}+1}\right)^{n}$ which is the distribution of a thermal state.
	
\begin{figure}[h]
        \centering
        \includegraphics[width=0.99\linewidth]{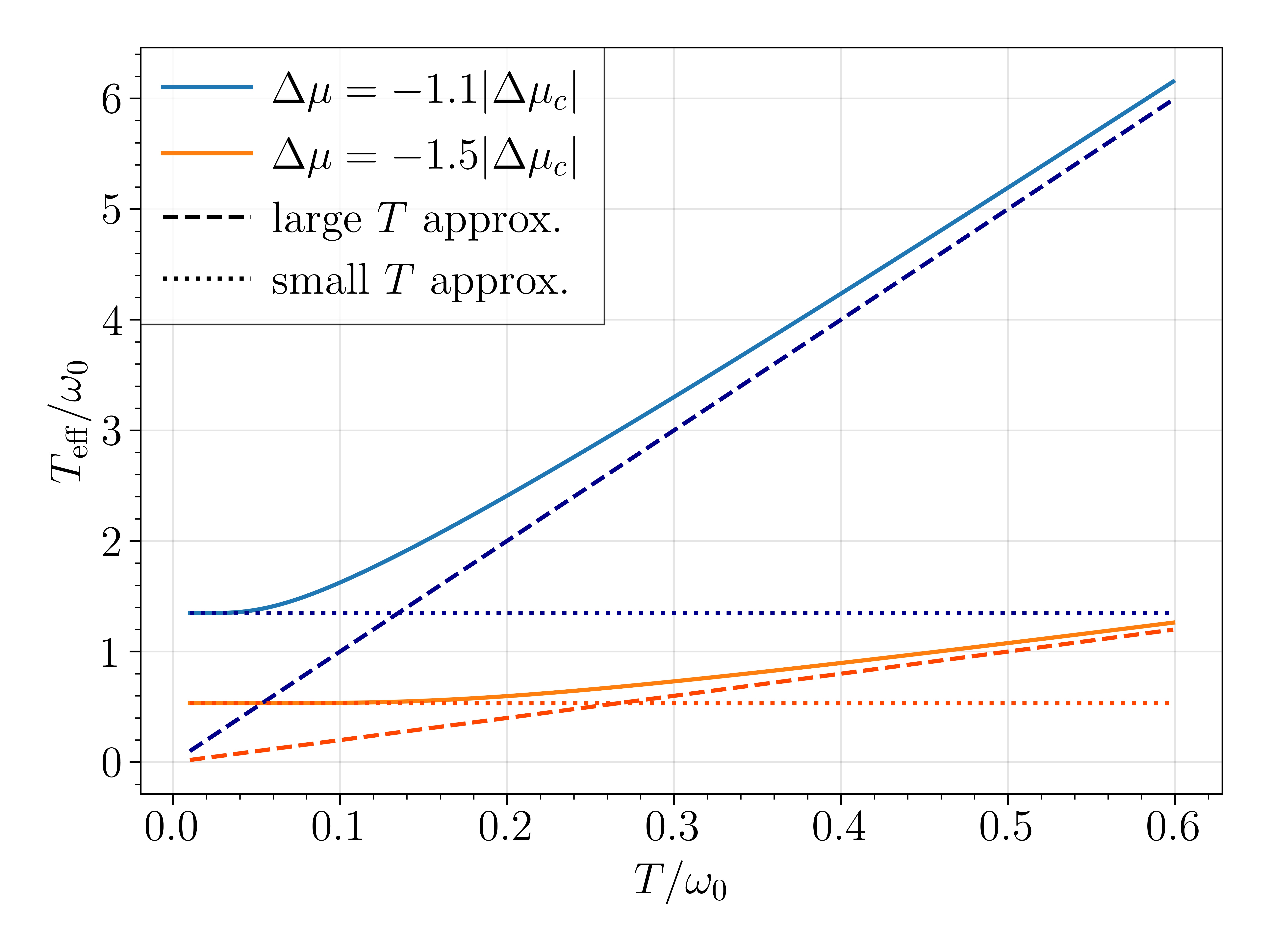}
        \caption{We plot the effective temperature $T_{\mathrm{eff}} = 1/k_{\mathrm{B}}\beta_{\mathrm{eff}}$ [Eq.~\eqref{eq:beta_eff}] as a function of $T=1/k_B\beta$, with $T_p=T_{\mathrm{th}} = T$, for two values of spin accumulation $\Delta\mu = -1.1|\Delta\mu_c|$ (blue) and $\Delta\mu = -1.1|\Delta\mu_c|$ (orange), with $\Delta\mu_c/\omega_0 = (\alpha+\alpha_p)/\alpha_p$ and $\alpha=0.001$ and $\alpha_p=0.01$. We compare the full expression [Eq.~\eqref{eq:beta_eff}] with the small $T$ (dotted) [Eq.~\eqref{eq:beta_eff_small_T}]} and large $T$ (dashed) [Eq.~\eqref{eq:beta_eff_large}] approximations.
        \label{fig:Fig4}
    \end{figure}

	The total master equation for $p_n$ is obtained by adding the RHS of Eq.~\eqref{eq:bulk_bath} and the RHS of Eq.~\eqref{eq:SOT_bath}. We find the solution 
	\begin{align}
		p_{n} & =p_{0}\left(\frac{\gamma_{\text{th}}\left(n_{\text{th}}+1\right)+\gamma_{p}n_{p}}{\gamma_{\text{th}}n_{\text{th}}+\gamma_{p}\left(n_{p}+1\right)}\right)^{n}, \label{eq:p_n}
	\end{align}
	where the normalization is given by $p_{0}=\left(\bar{n}+1\right)^{-1}$ and the solution is stable if $\Delta\mu< -\left(\alpha+\alpha_{p}\right)\omega_0/\alpha_{p}$. This stability condition agrees with the results from Ref. \cite{harms_antimagnonics_2024} and our classical results in Sec.~\ref{sec:classical}. The occupation $\bar{n} =\sum_{n}np_{n}$ reads $\bar{n}=n_{\text{bulk}}+n_{\text{pump}}$ with 
	\begin{align}
		n_{\text{bulk}} & =-\left(n_{\text{th}}+1\right)\frac{\alpha\omega_0}{\left(\alpha+\alpha_{p}\right)\omega_0+\alpha_{p}\Delta\mu},\\
		n_{\text{pump}} & =n_{p}\frac{\alpha_{p}\left(\omega_0+\Delta\mu\right)}{\left(\alpha+\alpha_{p}\right)\omega_0+\alpha_{p}\Delta\mu}.
	\end{align}
    Here, $n_\mathrm{pump}$ denotes the magnon occupation due to SOT and is hence proportional to $\alpha_p$, whereas $n_\mathrm{bulk}$ is the contribution to the magnon occupation from the bulk damping which is proportional to $\alpha$. At the instability, the average magnon number scales as $\bar{n}\propto |\Delta\mu-\Delta\mu_c|^{\delta}$, where $\delta$ is a critical exponent that is given by $\delta=-1$. This result differs from the logarithmic critical behavior obtained in Sec.~\ref{sec:classical_results}. Here, we consider a single-mode system, whereas before, we considered a multimode two-dimensional system in Sec.~\ref{sec:classical}. The distribution $p_n$ [Eq.~\eqref{eq:p_n}] can be re-expressed as a thermal distribution with an effective inverse temperature $\beta_{\text{eff}}=1/k_{\text{B}}T_{\text{eff}}$ given by 
	\begin{align}
		\beta_{\text{eff}} & =\omega_0\ln\left(\frac{\gamma_{\text{th}}n_{\text{th}}-\gamma_{p}\left(n_{p}+1\right)}{\gamma_{\text{th}}\left(n_{\text{th}}+1\right)-\gamma_{p}n_{p}}\right). \label{eq:beta_eff}
	\end{align}
	The effective temperature $\beta_{\text{eff}}$ [Eq.~\eqref{eq:beta_eff}] can be used as a measure to quantify the fluctuations in our system. Assuming the bulk and SOT bath to have the same temperature $T_p = T_{\mathrm{th}}\equiv T$, in the limit of large $T$, we find that the effective temperature is linear in $T$, given by \begin{equation}
	    \beta_{\text{eff}} \approx-\left(1+\frac{\alpha_{p}}{\alpha+\alpha_{p}}\frac{\Delta\mu}{\omega_0}\right)\beta, \label{eq:beta_eff_large}
	\end{equation}
    with $\beta=1/k_{\text{B}}T$. In the limit of small $T$, the inverse effective temperature $\beta_\mathrm{eff}$ [Eq.~\eqref{eq:beta_eff}] becomes
	\begin{align}
		\beta_{\text{eff}} & \approx\frac{1}{\omega_0}\ln\left(-\frac{\alpha_{p}\left(\omega_0+\Delta\mu\right)}{\alpha\omega_0}\right). \label{eq:beta_eff_small_T}
	\end{align}
    Eq.~\eqref{eq:beta_eff_small_T} demonstrates that even in the limit of small $T$, the system has a finite effective temperature $T_{\mathrm{eff}}$, hence larger fluctuations than the vacuum. In Fig.~\ref{fig:Fig4}, we plot the effective temperature $T_{\mathrm{eff}}$ as a function of $T$ for two values of spin accumulation $\Delta\mu$, demonstrating that larger $|\Delta\mu|$ suppresses fluctuations. As one can see in Fig.~\ref{fig:Fig2}, the suppression of fluctuations is also a result following from the classical dynamics analyzed in Sec.~\ref{sec:classical}. Hence, this behavior is expected.
    
    The steady state distribution $p_n$ [Eq.~\eqref{eq:p_n}] can be probed via dispersive readout with a qubit~\cite{schuster2007, lachance-quirion2017}. We consider a spin qubit with level splitting $\omega_q$, represented by Pauli matrix $\hat{\sigma}_z$, interacting with the nanomagnet via interfacial exchange coupling in the far detuned limit. We find a direct dispersive interaction between the excitations on the inverted state and the qubit described by the Hamiltonian $\hat{\mathcal{H}}_I = -\chi\hat{\psi}^\dagger \hat{\psi} \hat{\sigma}_z$~\cite{romling2023, kamra2017}. The coupling strength $\chi$ can be determined to be
	\begin{align}
		\chi & =\frac{J_{\text{int}}N_{\text{int}}\left|\phi\right|^{2}}{2N}.
	\end{align}
    where $J_\mathrm{int}$ is the interfacial exchange strength, $N_\mathrm{int}$ the number of interfacial sites, $N$ the total number of sites in the nanomagnet and $\left|\phi\right|^2$ the average wave function of the qubit at the interface. The full Hamiltonian describing the coupled magnet-qubit system $\hat{\mathcal{H}}_\mathrm{qm}$ can be written as
    \begin{equation}
        \hat{\mathcal{H}}_\mathrm{mq} = -\omega_0\hat{\psi}^\dagger\hat{\psi} +\frac{1}{2}\left(\omega_q -2\chi\hat{\psi}^\dagger\hat{\psi}\right)\hat{\sigma}_z.\label{eq:H_mq}
    \end{equation}
    Note here that $\hat{\mathcal{H}}_\mathrm{qm}$ [Eq.~\eqref{eq:H_mq}] is diagonal, such that its eigenstates can be denoted as the product states $\ket{n,\sigma} = \ket{n}\ket{\sigma}$, where $\ket{n}$ represents a Fock state with $n$ antimagnonic excitations and $\ket{\sigma}$ denotes the qubit state that can be in the ground (excited) state $\ket{g}$ ($\ket{e}$). Because of the dispersive interaction, the qubit acquires a modification in its excitation frequency that depends on the magnetic state $\omega_{q,n} = \omega_q - 2\chi n$~\cite{schuster2007}. The frequency $\omega_{q,n}$ corresponds to the energy difference between states $\ket{n,g}$ and $\ket{n,e}$. 

\begin{figure}[h]
        \centering
        \includegraphics[width=0.99\linewidth]{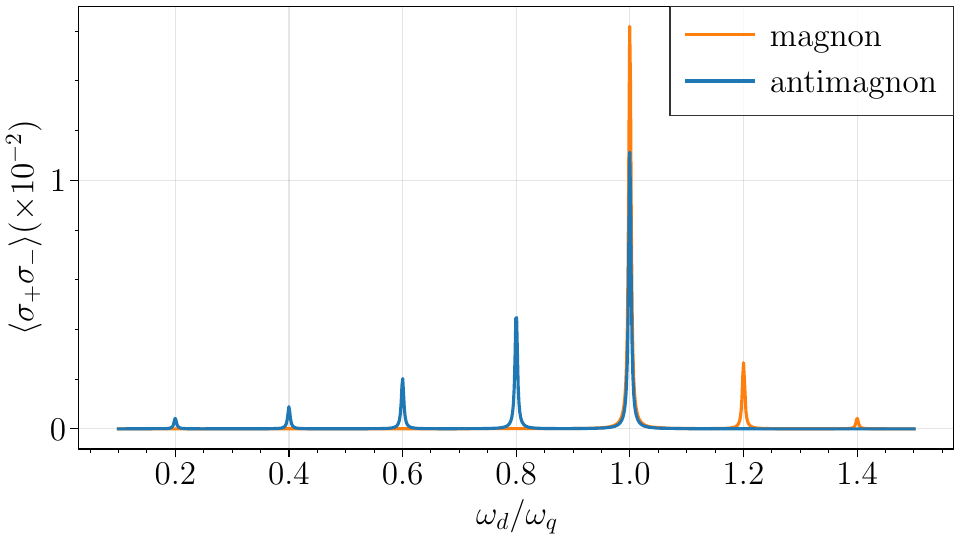}
        \caption{The steady state qubit excitation $\langle\hat{\sigma}_+ \hat{\sigma}_-\rangle$ is plotted as a function of the qubit drive frequency $\omega_d$. The qubit is coupled to a magnonic thermal state with temperature $k_\mathrm{B}T/\omega_0 = 0.55$ (orange) and an antimagnonic state with $k_\mathrm{B}T/\omega_0 = 0.55$, $\alpha=0.001$, $\alpha_p=0.1$,$\Delta\mu/\omega_0 = -1.5|\Delta\mu_c|$ (blue). For the spectroscopy, we employ the qubit frequency $\omega_q = 10\omega_0$, qubit dissipation $\gamma_q = 0.005\omega_q$, Rabi frequency $\Omega_d = \gamma/7$ and $\chi = 0.1\omega_q$.}
        \label{fig:Fig5}
    \end{figure}

    We employ a qubit drive modeled by the Hamiltonian $\hat{\mathcal{H}}_d=\Omega_d\cos(\omega_d)(\hat{\sigma}_+ + \hat{\sigma}_-)$ with Rabi frequency $\Omega_d$ and drive frequency $\omega_d$. The magnet is in the Fock state superposition $\ket{\Psi} = \sum_n p_n \ket{n}$. If the qubit is driven at frequency $\omega_d=\omega_{q,n}$, the transition $\ket{n,g}\rightarrow\ket{n,e}$ occurs with a probability of $|p_n|^2$. Sweeping over a range of frequencies and measuring the steady state qubit excitation $\langle \hat{\sigma}_+\hat{\sigma}_-\rangle$ reveals peaks in around frequencies $\omega_{q,n}$ with peak height proportional to the transition probability $|p_n|^2$. This way, the state of the magnet can be resolved. In Fig.~\ref{fig:Fig5}, we compare the qubit spectrum when coupled to the inverted state with the qubit spectrum when coupled to a thermal magnonic state. The magnonic Hamiltonian can be obtained from $\hat{\mathcal{H}}_\mathrm{mq}$ [Eq.~\eqref{eq:H_mq}] upon substitution $-\omega_0 \rightarrow \omega_0$ and $-\chi \rightarrow \chi$. Both states have the same input temperature $T$; however, due to SOT pumping, the antimagnonic state has more fluctuations, and, hence, a larger effective temperature in the distribution. Another feature is that antimagnonic peaks occur around frequencies $\omega_{q,n} = \omega_q - 2\chi n$, whereas magnonic peaks are mirrored and occur around $\omega_{q,n}^m = \omega_q + 2\chi n$.
    

	\section{Discussion and conclusions}\label{sec:Discussion and conclusions}
	In this article, we examined the fluctuations of the inverted magnetic state classically and quantum mechanically using the stochastic LLG and the master equation approach, respectively.
    The classical fluctuations are modelled by the stochastic fields in the bulk and at the interface.
    The bulk fluctuations are due to the scattering of electrons and phonons.
    The interfacial fluctuations are due to Joule heating and shot noise of the electron current, which is modelled using a stochastic field at the interface.
    A key finding is that fluctuations due to SOT play a crucial role in very thin ferromagnetic films ($d/\xi=0.001$), enhancing the magnon density by a factor of $\approx100$ compared to the thin film ferromagnet without fluctuations from SOT. This effect decreases for thicker films; for instance, the magnon density for a thin film of thickness $d/\xi=0.095$ increases by $\approx 10$ when considering interfacial contributions. This shows that there are regimes in which it is necessary to take fluctuations due to SOT into account since they enhance and shape the nature of excitations of the inverted state. 
	
	Another key finding is that large spin accumulation $|\Delta\mu|$ suppresses fluctuations while the fluctuations around the critical value $\Delta\mu_c$ diverge. This can be seen as a driven phase transition, where the value of spin accumulation $\Delta\mu$ determines whether the inverted state and ground state are stabilised or destabilised. For a realistic experimental set-up, the critical current density $j_c$ is given by $j_c=2\alpha \mu_0 e dM_s H_{exp}/\hbar \theta_{SH}$ with $e$ the elementary charge, $\hbar$ Plancks constant, and $\theta_{SH}$ the spin Hall angle \cite{liu_spin-torque_2011,lee2013threshold,harms_antimagnonics_2024}. The critical current density depends on material properties and ranges from $10^6$ to $10^{10}A/m^2$ for $\mu_0H_{exp}=0.5T$. In Appendix~\ref{sec:appendix_values}, we provide a list of materials with estimated critical current. Most of these currents are realistic for achieving the inverted state in experiments; however, they operate close to the instability and thus exhibit large fluctuations. 
    Large fluctuations can be used to inject thermal magnons with a high effective temperature, which is of interest for studying unusual phase transitions~\cite{skalyo1969}. 
    On the contrary, for logical operations, it is necessary to reduce the fluctuations as much as possible so that information does not get lost.
    Reduced fluctuations may be achieved via a large $|\Delta\mu|$.
    Unfortunately, arbitrarily large spin accumulation is not feasible in experiments.
    However, if there is an upper bound for the fluctuations, we can establish a lower bound on the spin accumulation. 
    Specifically, to achieve fluctuations equivalent to the non-driven ferromagnet for some temperature in the classical limit, we find that $\Delta\mu\approx2\Delta\mu_c$, given the symmetry of the fluctuations around $\Delta\mu_c$ in Fig. \ref{fig:Fig2}.
    This condition is equal to a charge current in the heavy metal layer being twice the critical current.
    When this is the case, the magnetoresistance of the charge current is approximately $0.8\Delta\rho_1/sd \lambda_T^2$, which is a direct consequence of the fluctuations in the inverted state.
    Therefore, it may be used as an experimental verification of the results in this article.

   Recently, two experimental papers \cite{Kurebayashi_Barker_Yamazaki_Kushwaha_Stenning_Youel_Hou_Dion_Prestwood_Bauer_et_al._2026,Karadza_Wang_Kercher_Noel_Legrand_Schlitz_Gambardella_2026} have shown the feasibility of the inverted magnetic state, for which the theoretical results in this article may prove to be insightful \footnote{In Ref. \cite{Kurebayashi_Barker_Yamazaki_Kushwaha_Stenning_Youel_Hou_Dion_Prestwood_Bauer_et_al._2026}, the inverted magnetic state is compared to the Kaptiza pendulum. However, the effect of the Kaptiza pendulum is a result of a changing effective potential, which is different from the damping-like driving of a ferromagnet with spin-transfer torque or spin-orbit torque.}.

    For small systems and ultra-low temperatures, the quantum description predicts that the antimagnon thermal spectrum mimics an effective temperature that is larger than the actual temperature of the setup.
    This is the result of the negative energy character of the antimagnon.
    The system excites an antimagnon when it emits a quantum of energy, and it deexcites an antimagnon when it absorbs a quantum of energy.
    This is the opposite for magnons in an equilibrium magnet.
    The difference causes the effective temperature for antimagnons to be constant for low temperatures since the system may always emit a quantum of energy and therefore create an antimagnon, as is seen in Fig. \ref{fig:Fig4}.
    We also suggest that using a dispersive read-out with a qubit could show this experimentally.

	As a final remark, the ferromagnet becomes dissipationless when $\Delta\mu=\Delta\mu_c$.
    For this critical driving, the Gilbert damping is exactly opposed. Thus, fluctuations, which typically dissipate in time, accumulate.
    This causes the magnetisation of the ferromagnet to explore its full phase space, similar to an active particle.
    This demonstrates the potential of the inverted state as a platform with many interesting features. In conclusion, we examined and modelled the fluctuations in the inverted magnetic state.
    We have shown how the fluctuations are influenced by SOT and spin accumulations. 
    These findings allow for a more careful theoretical analysis of setups using the inverted magnetic state and estimates for future experiments.
    To ensure a more accurate model, future research could focus on domain wall formation or the implementation of the spin-Seebeck effect in the inverted magnetic state.

    \begin{acknowledgments}
        The authors would like to thank A. Kamra, J. S. Harms, and  H. Y. Yuan for their input. A. E. R. acknowledges that the project that gave rise to these results received the support of a fellowship from ``la Caixa'' Foundation (ID 100010434) with the fellowship code LCF/BQ/DI22/11940029. A. L. B. acknowledges that the project is in part funded by “Black holes on a chip” with project number OCENW.KLEIN.502 and “Fluid Spintronics” with project number VI.C.182.069. Both are financed by the Dutch Research Council (NWO).
    \end{acknowledgments}

    \appendix
    \section{Full expression of the integrand \label{sec:appendix_integrand}}
    Here, we provides the full expression for the average occupation number $\langle \psi^*(x,\boldsymbol{\rho},t) \psi(x,\boldsymbol{\rho},t) \rangle $ following from stochastic LLG as discussed in Sec.~\ref{sec:LLG}. While Eqs.~\eqref{eq:integrand1} and \eqref{eq:integrand2} in the small $\kappa d\ll1$ limit gives us intuition about the systems behavior, the full expression provides information about especially about the $x$-dependence of the integrand. It reads 
    \begin{widetext}
    \begin{align}
\langle \psi^{*}\!&\left(x,\boldsymbol{\rho},t\right)\psi\!\left(x,\boldsymbol{\rho},t\right)\rangle   =\int\frac{\text{d}q}{2\pi}\int\frac{\text{d}\omega}{2\pi}q\frac{1}{A^{2}\left|W\right|^{2}}\Bigg\{ A^{2}\left|\kappa\right|^{2}\left|\cosh\left(\kappa x-\kappa d\right)\right|^{2}\frac{\frac{g^{\uparrow\downarrow}}{4\pi s}\left(\omega-\Delta\mu\right)}{\tanh\left(\frac{\omega-\Delta\mu}{2T}\right)}\nonumber \\
 & +\frac{\alpha\omega}{2\tanh\left(\frac{\omega}{2T}\right)}\Bigg[\left|\cosh\left(\kappa x-\kappa d\right)\right|^{2}\Bigg\{\left(\frac{g^{\uparrow\downarrow}}{4\pi s}\right)^{2}\left(\omega-\Delta\mu\right)^{2}\left(\frac{\sinh\left(\kappa x+\kappa^{*}x\right)}{\kappa+\kappa^{*}}-\frac{\sinh\left(\kappa x-\kappa^{*}x\right)}{\kappa-\kappa^{*}}\right)\nonumber \\
 & +A^{2}\left|\kappa\right|^{2}\left(\frac{\sinh\left(\kappa x+\kappa^{*}x\right)}{\kappa+\kappa^{*}}+\frac{\sinh\left(\kappa x-\kappa^{*}x\right)}{\kappa-\kappa^{*}}\right) +i\frac{g^{\uparrow\downarrow}}{4\pi s}\left(\omega-\Delta\mu\right)A\kappa\left(\frac{\cosh\left(\kappa x+\kappa^{*}x\right)-1}{\kappa+\kappa^{*}}-\frac{\cosh\left(\kappa x-\kappa^{*}x\right)-1}{\kappa-\kappa^{*}}\right)\nonumber \\
 & -i\frac{g^{\uparrow\downarrow}}{4\pi s}\left(\omega-\Delta\mu\right)A\kappa^{*}\left(\frac{\cosh\left(\kappa x+\kappa^{*}x\right)-1}{\kappa+\kappa^{*}}+\frac{\cosh\left(\kappa x-\kappa^{*}x\right)-1}{\kappa-\kappa^{*}}\right)\Bigg\}\nonumber \\
 & -\left|i\frac{g^{\uparrow\downarrow}}{4\pi s}\left(\omega-\Delta\mu\right)\sinh\left(\kappa x\right)-A\kappa\cosh\left(\kappa x\right)\right|^{2}\left(\frac{\sinh\left(\kappa x+\kappa^{*}x-\kappa d-\kappa^{*}d\right)}{\kappa+\kappa^{*}}+\frac{\sinh\left(\kappa x-\kappa^{*}x-\kappa d+\kappa^{*}d\right)}{\kappa-\kappa^{*}}\right)\Bigg]\Bigg\}.\label{eq:full_n}
\end{align}
    \end{widetext}
    with $\kappa$ from Eq.~\eqref{eq:kappa} the Wronskian $W$ from Eq.~\eqref{eq:wronskian}. 

    In the small $d\kappa\ll1$ limit, the $\omega$-integration for the integrands $I_p(\omega,\boldsymbol{q})$ and $I_{\text{th}}(\omega,\boldsymbol{q})$ [Eqs.~\eqref{eq:integrand1} and \eqref{eq:integrand2}] can be performed analytically using contour integration. We find the following expression
    \begin{widetext}
        \begin{align}
\langle \psi^{*}&\left(x,\boldsymbol{\rho},t\right)\psi\left(x,\boldsymbol{\rho},t\right)\rangle   =\nonumber \\
&=\frac{1}{d}\int\frac{\text{d}^{2}\boldsymbol{q}}{\left(2\pi\right)^{2}} \frac{\alpha_p\left(\omega_0-A\boldsymbol{q}^{2}+\Delta\mu\right)\left(n_{B}^p\left(-\left(\omega_0-A\boldsymbol{q}^{2}+\Delta\mu\right)\right)+\frac{1}{2}\right)-\alpha\left(\omega_0-A\boldsymbol{q}^{2}\right)\left(n_{B}^\mathrm{th}\left(\omega_0-A\boldsymbol{q}^{2}\right)+\frac{1}{2}\right)}{\alpha_p\left(\omega_0-A\boldsymbol{q}^{2}+\Delta\mu\right)+\alpha\left(\omega_0-A\boldsymbol{q}^{2}\right)}.
\end{align}
    \end{widetext}

    \section{Estimation of experimental values \label{sec:appendix_values}}

In this Appendix, we expand on the values for the critical current to achieve the inverted magnetic state.
We consider three heavy metals to inject SOT, $Pt$, $Ta$, and $W$.
They respectively have the spin Hall angles, $0.08$\cite{SHA:Pt}, $0.15$\cite{SHA:Ta}, and $0.3$\cite{SHA:W}.
For the ferromagnets, we investigate Permalloy, CoFeB, and YIG.
They have a saturation magnetization and Gilbert damping, which are respectively given by $\mu_0M_S=0.85T,1.32 T,0.16T$ and $\alpha=0.007,0.004,10^{-5}$, in Ref.\cite{Permalloy,CoFeB,YIG}.
These values allow us to compute table \ref{tab:current_density} with critical current densities for an external field of $\mu_0H=0.5T$.
The critical current densities are between the ranges $10^6A/m^2$ and $10^{10}A/m^2$.

\begin{table}[h]
    \centering
    \begin{tabular}{|c||c||c||c|}
& Pt & Ta & W\\ \hline
Permalloy & $1.43\times 10^{10}$ & $7.62\times 10^9$ & $3.81\times 10^9$ \\
CoFeB & $1.27\times 10^{10}$ & $6.76\times 10^9$ & $3.38\times 10^9$ \\
YIG & $3.89\times 10^6$ & $2.08\times 10^6$ & $1.04\times 10^6$
    \end{tabular}
    \caption{Table of critical current densities in heavy metals (horizontal) necessary to invert the magnetization in the FMs (vertical). The current density has units $A/m^2$.}
    \label{tab:current_density}
\end{table}

\bibliography{BIB}

\end{document}